\documentclass[aps,preprint,superscriptaddress,nofootinbib,preprintnumbers,eqsecnum,prd]{revtex4-1}

\usepackage{bm, amsmath,amssymb,amsfonts,slashed,array,graphicx,soul,multirow,mathtools}
\usepackage[vcentermath,noautoscale]{youngtab}

\Yboxdim{8pt}

\usepackage{xfrac}

\def\lsim{\mathrel{\rlap{\lower4pt\hbox{\hskip1pt$\sim$}}
    \raise1pt\hbox{$<$}}}
\def\gsim{\mathrel{\rlap{\lower4pt\hbox{\hskip1pt$\sim$}}
    \raise1pt\hbox{$>$}}}

\usepackage{xcolor}
\definecolor{red}{rgb}{0.9, 0,0}

\usepackage{hyperref}	
\usepackage{url,comment}
\usepackage{times}
\usepackage{dblfloatfix}
\usepackage{latexsym}
\usepackage{graphicx, graphics, hyperref, amsmath, amssymb, slashed, xcolor, bbm,bm,amsthm, array}
 \usepackage{subfigure}
 \usepackage{listings} 
 \usepackage{wrapfig}
 \usepackage{trimclip}

\usepackage{rotating}
\usepackage{afterpage}
 
\newcommand{\nc}{\newcommand}

\nc{\beq}{\begin{equation}}
\nc{\eeq}{\end{equation}}
\nc{\barray}{\begin{eqnarray}}
\nc{\earray}{\end{eqnarray}}
\nc{\barrayn}{\begin{eqnarray*}}
\nc{\earrayn}{\end{eqnarray*}}
\nc{\bcenter}{\begin{center}}
\nc{\ecenter}{\end{center}}
\nc{\mc}{\mathcal}
\nc{\er}[1]{(\ref{eq:#1})}
\nc{\onehalf}{\frac{1}{2}} 
\nc{\partialbar}{\bar{\partial}}
\nc{\psit}{\widetilde{\psi}}
\nc{\Tr}{\mbox{Tr}}
\nc{\hc}{\mbox{H.c.}}
\nc{\ev}{\;\mathrm{eV}}
\nc{\mev}{\;\mathrm{MeV}}
\nc{\gev}{\;\mathrm{GeV}}
\nc{\kev}{\;\mathrm{keV}}
\nc{\tev}{\;\mathrm{TeV}}
\nc{\pev}{\;\mathrm{PeV}}
\nc{\eev}{\;\mathrm{EeV}}

\def\chii0{\chi_i^0}
\def\chij0{\chi_j^0}

\nc{\ttbar}{t\bar t}

\newcommand{\cref}[1]{Chapter~\ref{c.#1}}

\begin{document}

\title{\thispagestyle{empty}
MATHUSLA: 
\\ A Detector Proposal to Explore
\\
the Lifetime Frontier at the HL-LHC
\\ \vspace{0.5cm}
\normalsize\normalfont{Input to the update process of the European Strategy for Particle Physics}
\\
\normalsize\normalfont{18. December 2018}
}

\author{Henry Lubatti (Corresponding Author)}
\email{lubatti@uw.edu}
\affiliation{University of Washington, Seattle}

\author{Cristiano Alpigiani}
\affiliation{University of Washington, Seattle}

\author{Juan Carlos Arteaga-Vel\'azquez}
\affiliation{Universidad Michoacana de San Nicol\'as de Hidalgo, Mexico (UMSNH)}

\author{Austin Ball}
\affiliation{CERN}

\author{Liron Barak}
\affiliation{Tel Aviv University}

\author{James Beacham}
\affiliation{Ohio State University}

\author{Yan Benhammo}
\affiliation{Tel Aviv University}

\author{Karen Salom\'e Caballero-Mora}
\affiliation{Universidad Aut\'onoma de Chiapas, Mexico (UNACH)}

\author{Paolo Camarri}
\affiliation{Sezione di Roma Tor Vergata, Roma, Italy}

\author{Tingting Cao}
\affiliation{Tel Aviv University}

\author{Roberto Cardarelli}
\affiliation{Sezione di Roma Tor Vergata, Roma, Italy}

\author{John Paul Chou}
\affiliation{Rutgers, the State University of New Jersey}

\author{David Curtin}
\affiliation{University of Toronto}

\author{Albert de Roeck}
\affiliation{CERN}

\author{Giuseppe Di Sciascio}
\affiliation{Sezione di Roma Tor Vergata, Roma, Italy}

\author{Miriam Diamond}
\affiliation{University of Toronto}

\author{Marco Drewes}
\affiliation{Universit\'{e} catholique de Louvain}

\author{Sarah C. Eno}
\affiliation{University of Maryland}

\author{Rouven Essig}
\affiliation{YITP Stony Brook}

\author{Jared Evans}
\affiliation{University of Cincinnati}

\author{Erez Etzion}
\affiliation{Tel Aviv University}

\author{Arturo Fern\'andez T\'ellez}
\affiliation{Benem\'erita Universidad Aut\'onoma de Puebla, Mexico (BUAP)}

\author{Oliver Fischer}
\affiliation{Karlsruhe Institute of Technology}

\author{Jim Freeman}
\affiliation{Fermi National Accelerator Laboratory (FNAL)}

\author{Stefano Giagu}
\affiliation{Universit\`{a} degli Studi di Roma La Sapienza, Roma, Italy}

\author{Brandon Gomes}
\affiliation{Rutgers, the State University of New Jersey}

\author{Andy Haas}
\affiliation{New York University}

\author{Yuekun Heng}
\affiliation{Institute of High Energy Physics, Beijing}

\author{Giuseppe Iaselli}
\affiliation{Politecnico di Bari, Italy}

\author{Ken Johns}
\affiliation{University of Arizona}

\author{Muge Karagoz}
\affiliation{University of Maryland}

\author{Audrey Kvam}
\affiliation{University of Washington, Seattle}

\author{Dragoslav Lazic}
\affiliation{Boston University}

\author{Liang Li}
\affiliation{Shanghai Jiao Tong University}

\author{Barbara Liberti}
\affiliation{Sezione di Roma Tor Vergata, Roma, Italy}

\author{Zhen Liu}
\affiliation{University of Maryland}

\author{Giovanni Marsella}
\affiliation{Universit\`{a} del Salento, Lecce, Italy}

\author{Piter A. Paye Mamani}
\affiliation{Instituto de Investigaciones F\'isicas (IIF), Observatorio de F\'isica C\'osmica de âChacaltayaâ, Universidad Mayor de San Andr\'es (UMSA)}

\author{Mario Iv\'{a}n Mart\'inez Hern\'andez}
\affiliation{Benem\'erita Universidad Aut\'onoma de Puebla, Mexico (BUAP)}

\author{Matthew McCullough}
\affiliation{CERN}

\author{David McKeen}
\affiliation{TRIUMF}

\author{Patrick Meade}
\affiliation{YITP Stony Brook}

\author{Gilad Mizrachi}
\affiliation{Tel Aviv University}

\author{David Morrissey}
\affiliation{TRIUMF}

\author{Meny Raviv Moshe}
\affiliation{Tel Aviv University}

\author{Antonio Policicchio}
\affiliation{Universit\`{a} degli Studi di Roma La Sapienza, Roma, Italy}

\author{Mason Proffitt}
\affiliation{University of Washington, Seattle}

\author{Marina Reggiani-Guzzo}
\affiliation{University of Campinas}

\author{Mario Rodr\'iguez-Cahuantzi}
\affiliation{Benem\'erita Universidad Aut\'onoma de Puebla, Mexico (BUAP)}

\author{Joe Rothberg}
\affiliation{University of Washington, Seattle}

\author{Rinaldo Santonico}
\affiliation{Sezione di Roma Tor Vergata, Roma, Italy}

\author{Marco Schioppa}
\affiliation{INFN and University of Calabria}

\author{Jessie Shelton}
\affiliation{University of Illinois Urbana-Champaign}

\author{Brian Shuve}
\affiliation{Harvey Mudd College}

\author{Yiftah Silver}
\affiliation{Tel Aviv University}

\author{Daniel Stolarski}
\affiliation{Carleton Unversity}

\author{Martin A. Subieta Vasquez}
\affiliation{Instituto de Investigaciones F\'isicas (IIF), Observatorio de F\'isica C\'osmica de âChacaltayaâ, Universidad Mayor de San Andr\'es (UMSA)}

\author{Guillermo Tejeda Mu\~{n}oz}
\affiliation{Benem\'erita Universidad Aut\'onoma de Puebla, Mexico (BUAP)}

\author{Steffie Ann Thayil}
\affiliation{Rutgers, the State University of New Jersey}

\author{Yuhsin Tsai}
\affiliation{University of Maryland}

\author{Emma Torro}
\affiliation{University of Washington, Seattle}

\author{Gordon Watts}
\affiliation{University of Washington, Seattle}

\author{Charles Young}
\affiliation{SLAC National Accelerator Laboratory}

\author{Jose Zurita}
\affiliation{Karlsruhe Institute of Technology Institute for Theoretical Physics}

\begin{abstract}

The observation of long-lived particles at the LHC would reveal physics beyond the Standard Model, could account for the many open issues in our understanding of our universe, and conceivably point to a more complete theory of the fundamental interactions. 
Such long-lived particle signatures are fundamentally motivated and can appear in virtually every theoretical construct that address the Hierarchy Problem, Dark Matter, Neutrino Masses and the Baryon Asymmetry of the Universe.  We describe in this document a large detector, MATHUSLA, located on the surface above an HL-LHC $pp$ interaction point, that could observe long-lived particles with lifetimes up to the Big Bang Nucleosynthesis limit of 0.1 s.  We also note that its large detector area allows MATHUSLA to make important contributions to cosmic ray physics.  Because of the potential for making a major breakthrough in our conceptual understanding of the universe, long-lived particle searches should have the highest level of priority.

\end{abstract}

\begin{center}
\includegraphics[width=5cm]{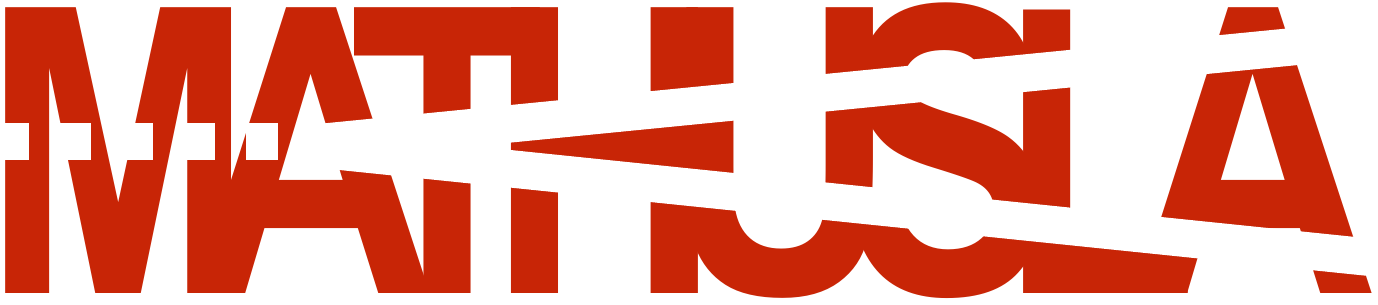}
\end{center}
\vspace{0.5cm}

\maketitle
\thispagestyle{empty}


\clearpage
\setcounter{page}{1}

\section{Introduction and Executive Summary}
\vspace{-3mm}

The Large Hadron Collider (LHC) has two main goals: determining the mechanism of electroweak symmetry breaking, and finding new physics beyond the Standard Model (BSM). 
While the discovery of the Higgs boson and the subsequent study of its detailed properties fulfill the first goal, convincing signs of BSM physics are yet to be found. 
The motivations for new physics are not diminished, with many fundamental mysteries begging for explanations outside of the Standard Model (SM). Where could the new physics at the LHC be hiding? MATHUSLA~\cite{Chou:2016lxi} is a proposal to address the significant gap in the LHC's reach for long-lived particles. \textbf{The LHC is the world-wide flag-ship experiment of particle physics and represents a huge investment. Augmenting its capabilities with relatively modestly-priced external detectors like MATHUSLA to maximize the discovery potential for new physics should be a high-priority goal.}

\begin{figure}
\begin{center}
\includegraphics[width=0.9 \textwidth]{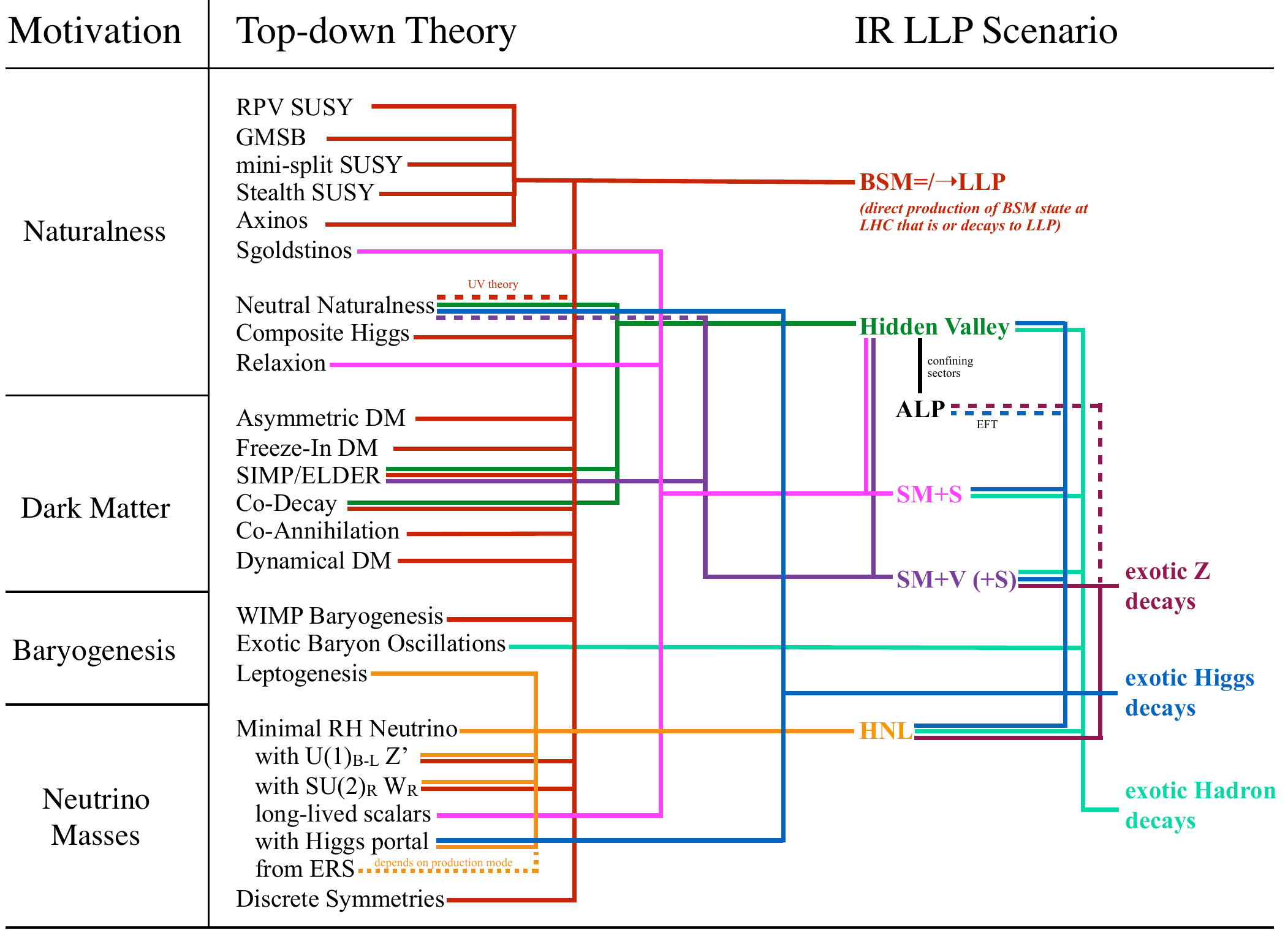}
\end{center}
\caption{
Summary of some top-down theoretical motivations for LLP signals at MATHUSLA~\cite{Curtin:2018mvb}.
}
\label{f.theorysummary}
\end{figure}

The existing BSM search programs, which mostly focus on energetic final states produced within subatomic distances of the proton collision, may have (literally) been looking in the wrong place. 
These searches are largely insensitive to neutral \emph{Long-Lived Particles} (LLPs), which are invisible until they decay into visible SM particles some macroscopic distance away from the the interaction point (IP). 
Far from being exotic oddities, LLP signatures are fundamentally motivated and could explain the Hierarchy Problem, Dark Matter, Neutrino Masses and the Baryon Asymmetry of the Universe~\cite{Curtin:2018mvb}, see Fig.~\ref{f.theorysummary}.
On more generic grounds, particles with macroscopic lifetimes are also ubiquitous in the SM, and the same mechanisms could be operating in any BSM theory. 
\textbf{In light of their fundamental motivation and LHC null results,  LLP searches should have the highest level of priority.}

An LLP decaying in LHC main detectors can be reconstructed as a \emph{displaced vertex} (DV). The geometrical nature of this signature makes it highly spectacular in many cases, and dedicated searches at ATLAS and CMS are ramping up to look for a variety of LLP signals. 
However, trigger and background limitations severely curtail the range of LLP masses, decay modes and lifetimes to which the LHC main detectors are sensitive.
Particularly challenging  are LLPs with \emph{very long lifetimes} that decay dominantly outside of the detector.
While low-energy and fixed-target experiments like SHiP, NA62, SEAQUEST, etc. may ~play an important role, many predicted LLPs can only be produced in the high-energy collisions of the LHC.
\textbf{The LHC could produce many LLPs with MeV - TeV masses that cannot be produced anywhere else, but that existing detectors cannot discover.}

MATHUSLA~\cite{Chou:2016lxi} is a proposal to address this extensive blind spot. The proposed large-scale surface detector located above CMS or ATLAS can detect LLPs with lifetimes near the cosmological limit of 0.1 s~\cite{Fradette:2017sdd}, and will extend the sensitivity of the main detectors by orders of magnitude for large classes of highly motivated LLP signatures.

As a secondary physics objective, MATHUSLA would also be able to perform cutting-edge cosmic ray physics observations to elucidate the nature of galactic cosmic rays, supernovae and other sources, and help solve important puzzles in astroparticle physics. This represents a guaranteed physics return on the investment of building the detector.

The MATHUSLA collaboration has already operated a test stand detector above ATLAS, made significant progress on detailed background and  design studies, and recently presented a Letter of Intent~\cite{Alpigiani:2018fgd} to the LHCC.
A significant fraction of the particle physics community came together to make the case for building MATHUSLA (and the importance of LLP searches in general) in a comprehensive white paper~\cite{Curtin:2018mvb}.
The collaboration is now seeking funding for the required R\&D to ensure this large-scale detector can be built at a reasonable cost, and for the construction of a MATHUSLA demonstrator detector unit by 2021. The full-scale detector could then become operational by 2025-26.

\section{Basic Detector Principles}
\vspace{-3mm}

\begin{figure}
\begin{center}
\begin{tabular}{m{0.5\textwidth}m{0.4\textwidth}}
\includegraphics[width=0.5\textwidth]{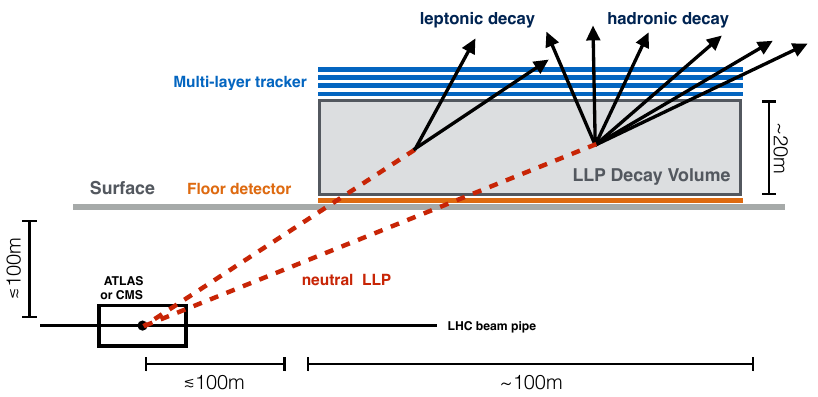}
&
\includegraphics[width=0.4\textwidth]{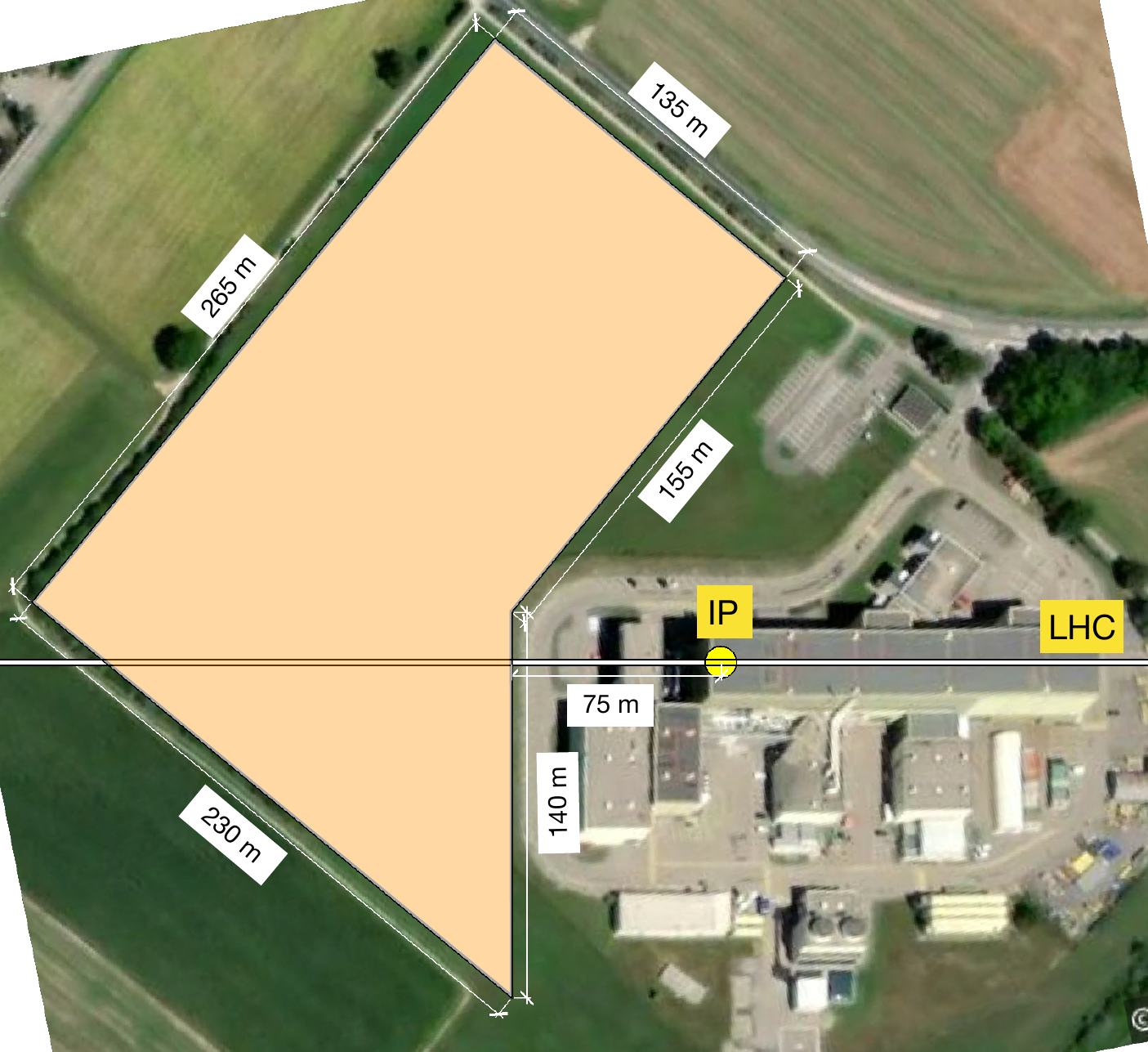}
\end{tabular}
\end{center}
\caption{
\emph{Left:} Simplified MATHUSLA detector layout with leptonic or hadronic LLP decay.  .
\emph{Right:} 
CERN-owned land near CMS (orange) that would be a suitable site for MATHUSLA. 
An optimized geometry on a fraction of this available land would achieve the same sensitivity as the original $200\mathrm{m} \times 200 \mathrm{m}$ benchmark~\cite{Chou:2016lxi, Curtin:2018mvb} while having only 
$\sim 1/3$
that area.
}
\label{f.mathuslalayout}
\end{figure}

MATHUSLA (MAssive Timing Hodoscope for Ultra-Stable neutraL pArticles)~\cite{Chou:2016lxi}  
is a proposed surface detector above ATLAS or CMS that can detect the decays of neutral LLPs in a low-background environment. The basic layout is shown in Fig.~\ref{f.mathuslalayout} (left). 
An empty air-filled fiducial decay volume is monitored by a robust multi-layer tracking system in the roof of the detector structure. (An excavated decay volume with the tracker at ground level is another option.) LLP decays are reconstructed as displaced vertices of upwards traveling charged particles. 

MATHUSLA's position on the surface,  separated from the LHC collision by $\sim 100 \mathrm{m}$ of rock, shields it from the ubiquitous QCD backgrounds that curtail the ability of the LHC main detectors to discover LLPs. 
To maintain reasonable geometric acceptance for LLPs ($\sim 5\%$ of solid angle), the detector must be very large, with linear dimensions of $\mathcal{O}(100 \mathrm{m})$ and a height of $\sim 20\mathrm{m}$.

A suitable site for the detector has been identified. CERN owns an available piece of land near CMS~\cite{Alpigiani:2018fgd}, see Fig.~\ref{f.mathuslalayout} (right). 
The original proposal~\cite{Chou:2016lxi} and the LOI~\cite{Alpigiani:2018fgd} defined a simplified square detector geometry with an area of $200\mathrm{m} \times 200\mathrm{m}$ and a height of $20\mathrm{m}$ for the decay volume, which was displaced from the IP by 100m both horizontally and vertically. 
The same ``MATHUSLA200'' geometry benchmark was also used to estimate the physics reach of the MATHUSLA detector for a large variety of LLP scenarios in the Physic Case white paper~\cite{Curtin:2018mvb}.
Since the available CMS site is closer to the IP both horizontally and vertically, a realistic geometry with $\sim 1/3$ the area of MATHUSLA200 can reach the same LLP sensitivity. %
For this reason we continue to use MATHUSLA200 as a physics reach benchmark throughout this and other documents, while emphasizing that the final detector design will reach this sensitivity with a more optimized, smaller and closer geometry that is tailored to the available experimental site.
This will also be an important factor in reducing the cost of the full detector.

For LLPs with lifetimes $\gtrsim 100\mathrm{m}$, MATHUSLA will have as many LLP decays in its detector volume as will ATLAS or CMS.
Crucial to its greater LLP sensitivity is the fact that unlike the main detectors, MATHUSLA can search for LLP decays without trigger restrictions and in the near-zero-background regime.

The dominant background on the surface is cosmic rays (CRs), which are incident on the full detector with a rate in the MHz range, corresponding to $\sim 10^{15}$ charged tracks over the whole HL-LHC run. 
Their rejection depends on the robust ceiling tracking system, comprised of $\sim$5 layers (the required number of layers will be determined by detailed study) with spatial and temporal resolutions in cm and nanosecond range, respectively. If the layers of this tracking system span a vertical distance of a few meters, full 4-dimensional track and displaced vertex reconstruction is possible, which significantly reduces the combinatorial backgrounds since associated tracks must intersect in both space and time to form a vertex.  This is an extremely stringent signal requirement even for LLPs with just two charged final states, but especially for hadronic LLP decays with $\mathcal{O}(10)$ charged final states.  Both Resistive Plate Chambers (RPCs) and plastic scintillators are time-tested technologies that easily meet the specifications needed for stringent background rejection. As argued in~\cite{Chou:2016lxi}, since CRs travel downwards and do not inherently form DVs, this signal requirement is expected to allow MATHUSLA to reach the near-zero-background regime.
 
 Other backgrounds are easier to handle. Upwards traveling muons from the LHC do not give a DV or, if they scatter or undergo rare decays that mimic LLP decays, can be vetoed by the floor detector. Neutrinos from atmospheric cosmic rays and the LHC scatter off air in the detector volume $\sim 100$ times during the entire HL-LHC run, but can be rejected with geometrical cuts and timing vetoes on non-relativistic charged tracks associated with the scattering event.

Even though MATHUSLA is basically just a large particle tracker without any energy or momentum measurement, it will still be able to measure many important properties of any LLP decays it observes~\cite{Curtin:2017izq}. Final state multiplicity would distinguish between leptonic and hadronic decay modes, while the geometry of the DV can be used to measure the LLP Lorentz boost event-by-event. It would even be possible to use MATHUSLA as a trigger for the main detector. Together with off-line correlated information from the main detector, this will allow the properties of any discoverd LLP like mass, production and decay mode to be determined.

The  proposed  detector  is  able  to  resolve  DVs  from  LLPs  with  masses below$\sim$ GeV ($\sim$10 MeV) for production in weak-scale processes ($B$-decays), giving MATHUSLA excellent sensitivity to low-mass LLP models as well as LLPs at the weak scale or above.

\begin{figure}[h!]
\begin{center}
\subfigure[]{\label{f.mathuslaL100Array:a}\includegraphics[height=6cm]{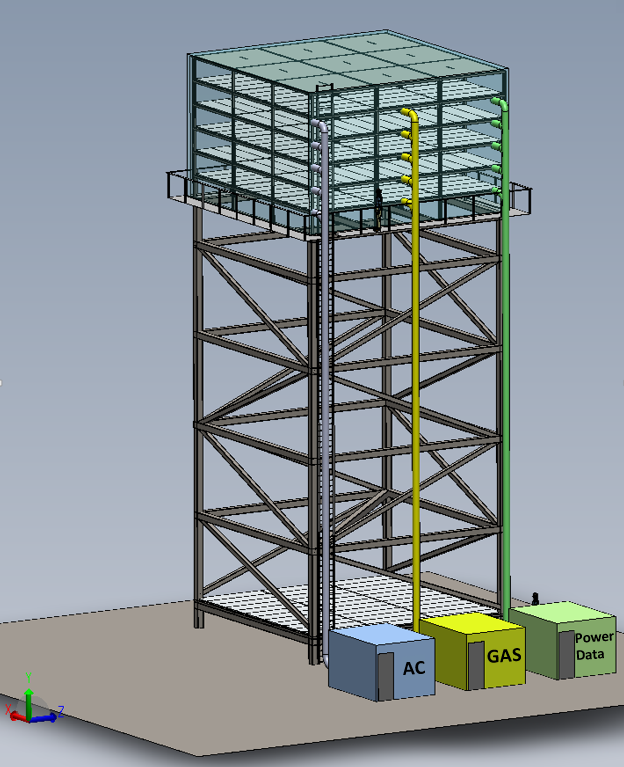}\label{f.mathuslaL100Array_a}}
\subfigure[]{\label{f.mathuslaL100Array:b}\includegraphics[height=6cm]{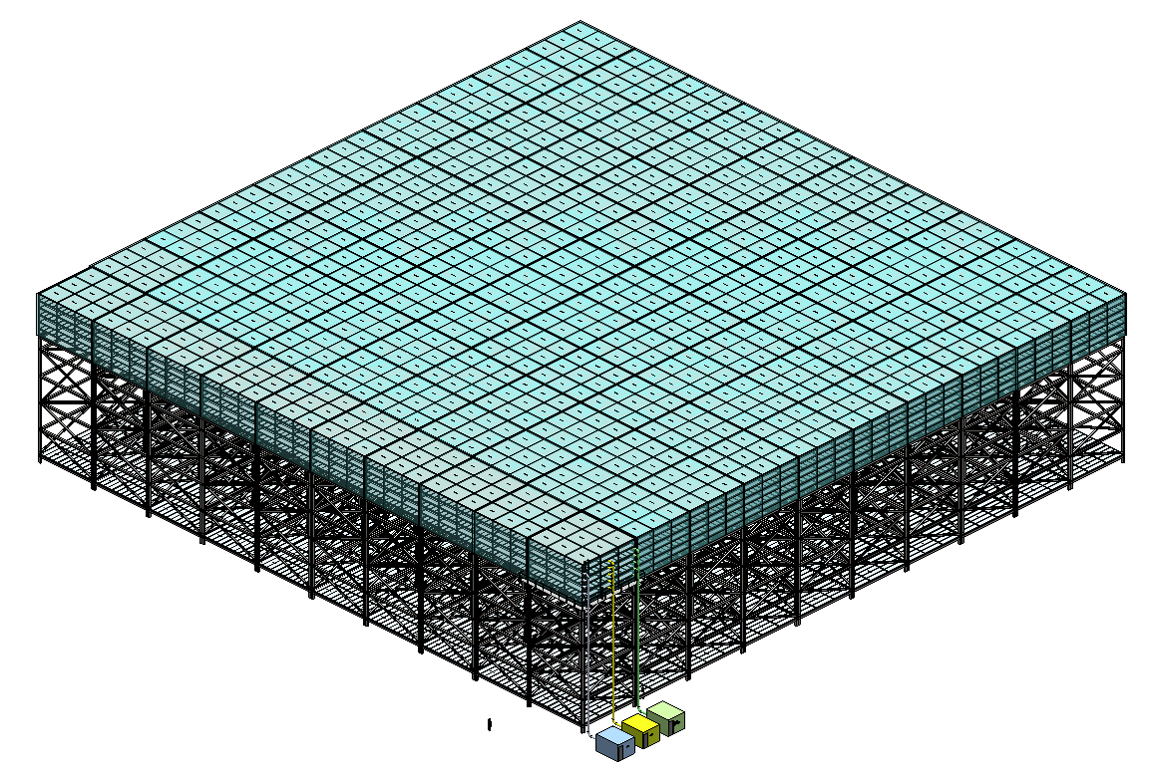}\label{f.mathuslaL100Array_b}}
\end{center}
\caption{
(a)~Conceptual design of one 10m $\times$ 10m MATHUSLA modular unit with stack of tracking layers on the top and one tracking layer at the floor. (b)~Conceptual design  an array of 100 MATHUSLA modular units. 
}
\label{f.mathuslaL100Array}
\end{figure}

While MATHUSLA leverages the  investments of the LHC and extends its physics reach, it is important to note that  MATHUSLA is entirely parasitic in nature, and its construction and operation are not expected to interfere with the operation of the LHC or its main experiments. 
MATHUSLA is also an inherently flexible detector concept that is scalable, lending itself to modular construction and staged implementation (see Fig.~\ref{f.mathuslaL100Array}). %

\section{The MATHUSLA Physics Case}
\vspace{-3mm}
The primary physics goal of MATHUSLA is the discovery of LLPs produced at the LHC, while cosmic ray physics represent an important secondary physics goal and a guaranteed return on the investment of building the detector.

\vspace{-3mm}
\subsection{Primary Physics Goal: Discovery of Long Lived Particles}
\vspace{-3mm}

\begin{figure}
\centering
\begin{tabular}{m{0.15\textwidth} m{0.45\textwidth}c m{0.3\textwidth}}
\includegraphics[width=0.15\textwidth]{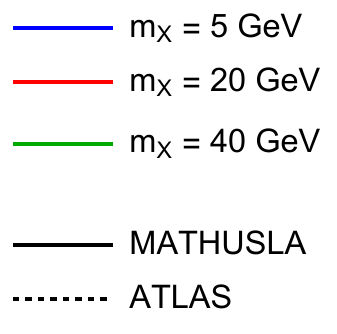}
&
\includegraphics[width=0.45\textwidth]{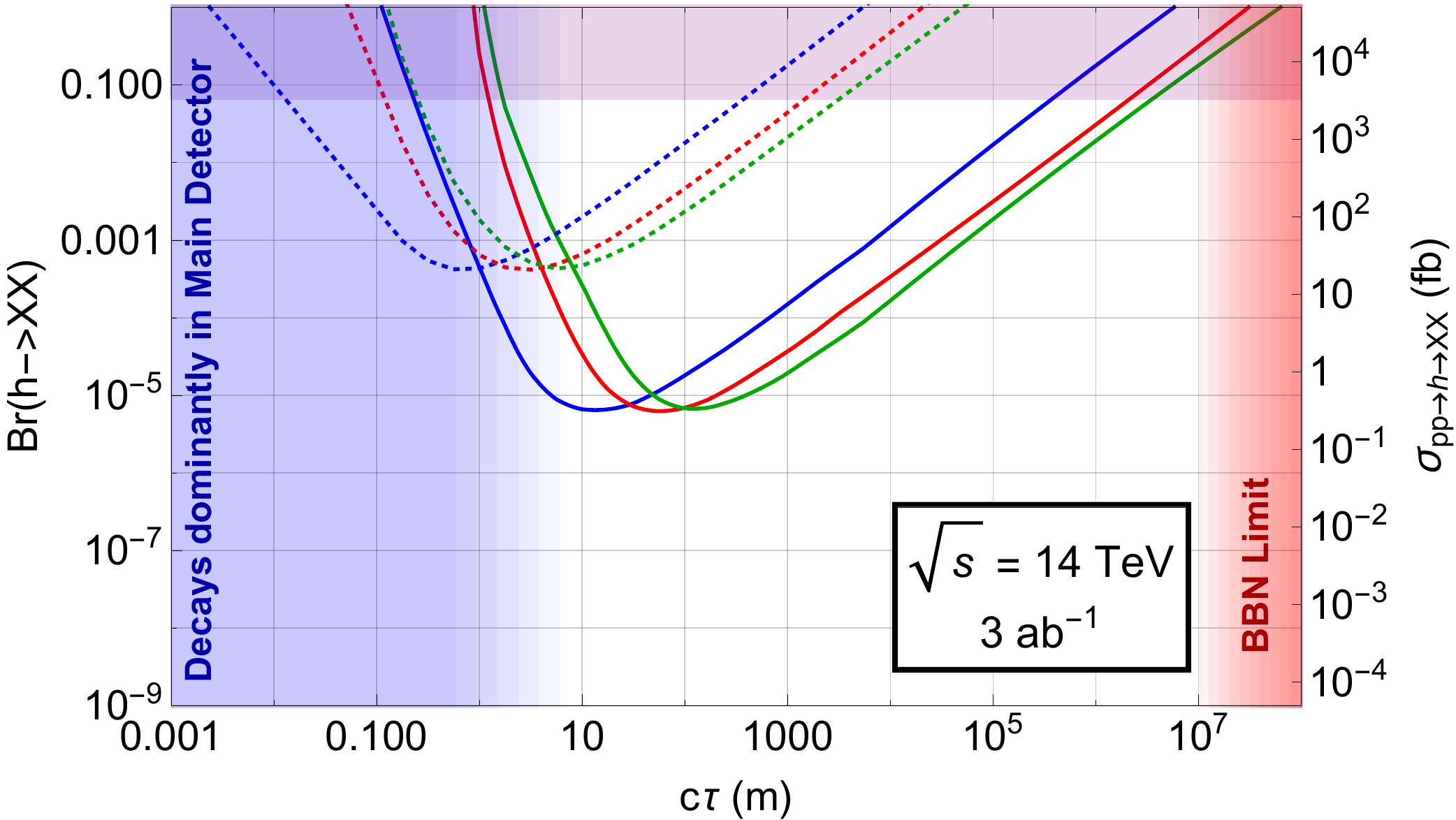}
&
\phantom{blabla}
&
\includegraphics[width=0.3\textwidth]{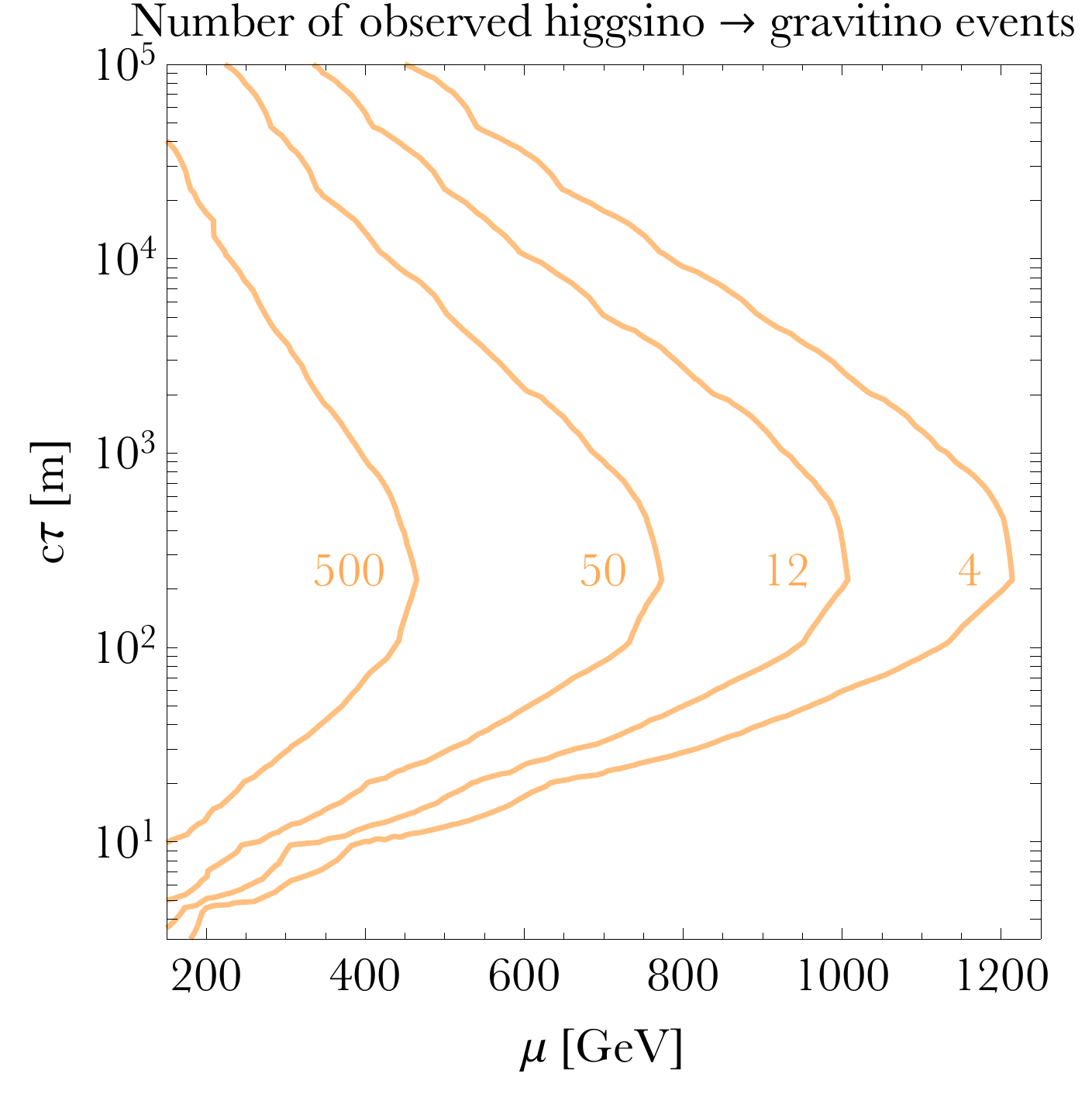}
\\
& \hspace{30mm} (a) && \hspace{25mm} (b)
\end{tabular}
\caption{MATHUSLA reach for weak- and TeV-scale LLP decays at the HL-LHC.
(a) Exotic Higgs decays to LLPs, MATHUSLA (solid) versus LHC main detectors (dashed)~\cite{Coccaro:2016lnz}. Figure from~\cite{Chou:2016lxi}.
(b) Long-lived Higgsinos in various models of supersymmetry. Contours show the number of Higgsino decays in MATHUSLA as a function of Higgsino mass $\mu$ and decay length $c\tau$. 
Plot reproduced from~\cite{Curtin:2018mvb}.
}
\label{f.MATHUSLAhiggssensitivity}
\end{figure}

\begin{figure}
\begin{center}
\begin{tabular}{m{0.4 \textwidth} c m{0.4 \textwidth} m{0.2 \textwidth} }
\includegraphics[width=0.4 \textwidth]{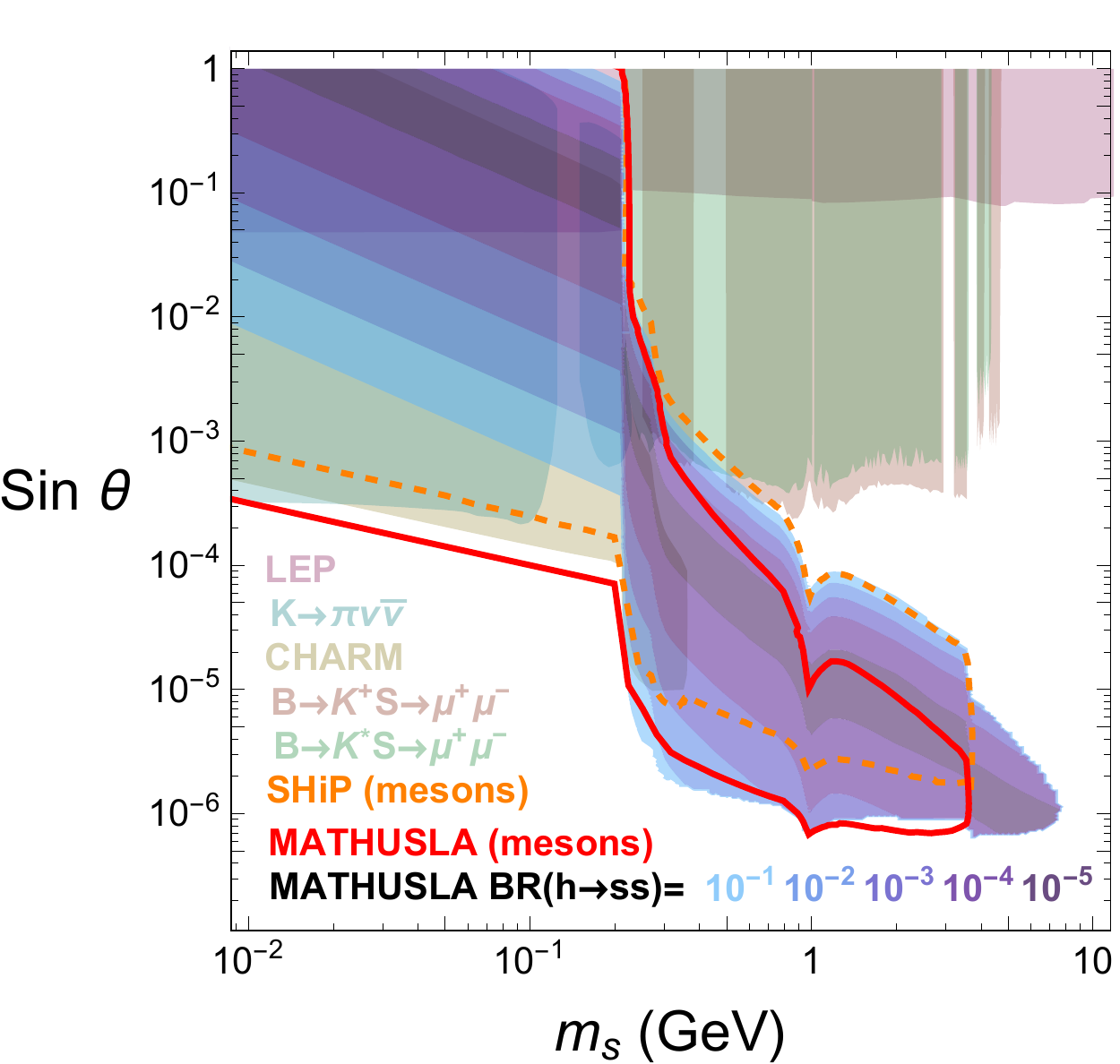}
&
\phantom{bla}
&
\includegraphics[width=0.4 \textwidth]{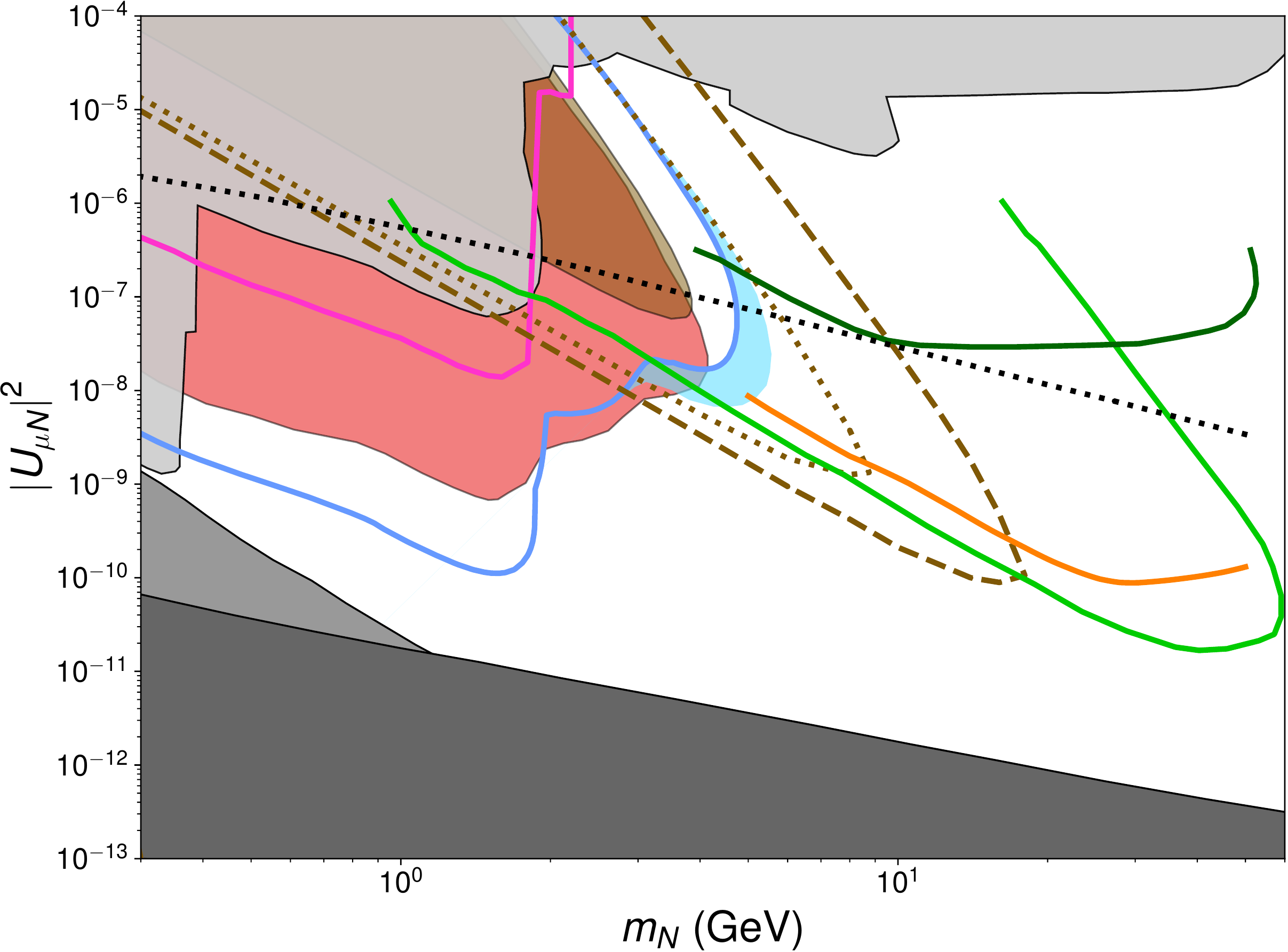}
&
\includegraphics[width=0.2 \textwidth]{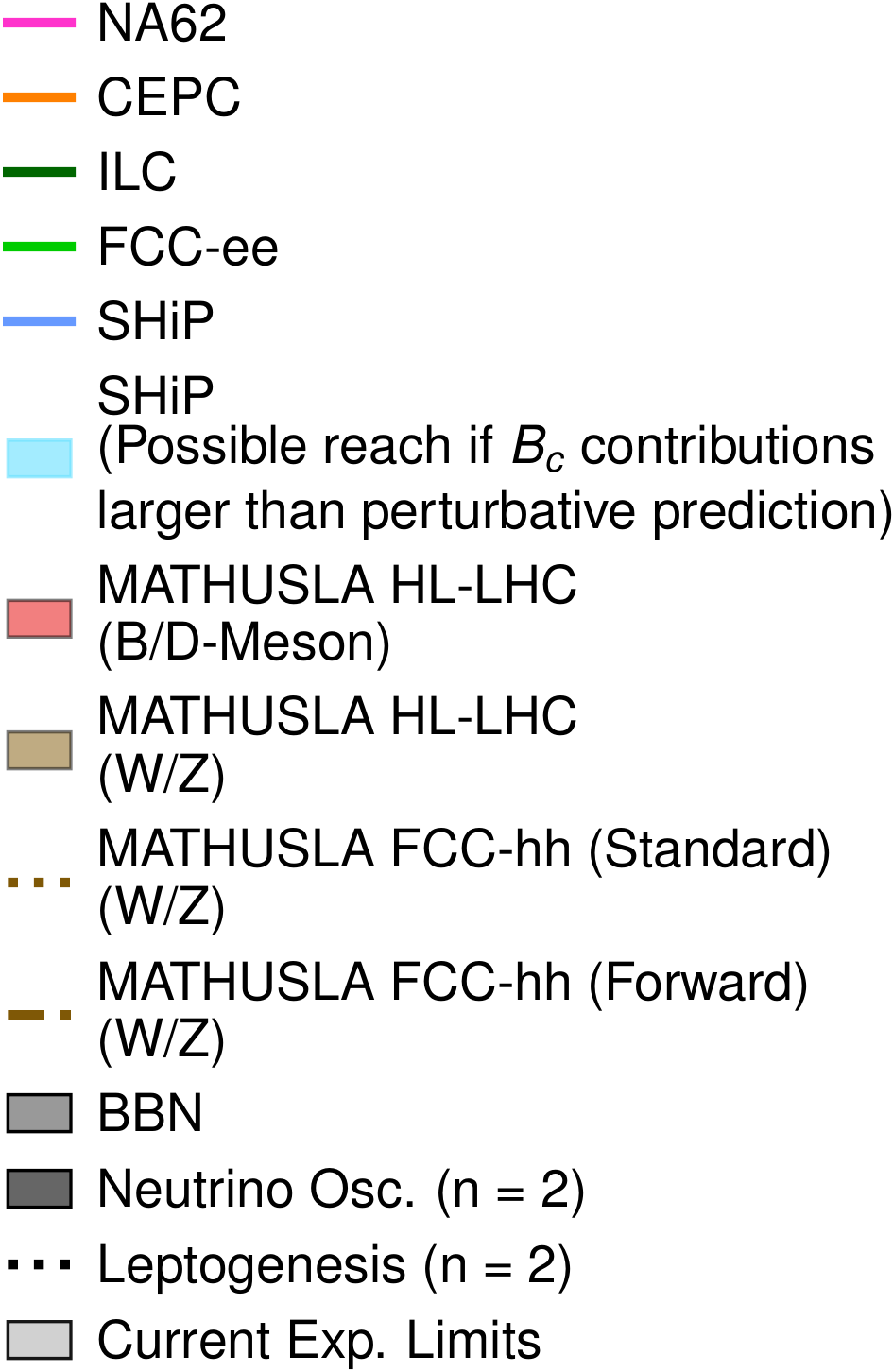}
\\
\hspace{0.2 \textwidth}  (a) && \hspace{0.2 \textwidth} (b)
\end{tabular}
\end{center}
\vspace{-4mm}
\caption{
Projected MATHUSLA sensitivity to low-mass LLPs at the HL-LHC. \emph{Note: these figures from~\cite{Curtin:2018mvb} are superseded by the official comparisons conducted within the PBC working group  and submitted separately to this European Strategy Update, but the relevant conclusions are unchanged.}
(a) ``Dark Scalar'' LLPs in the minimal SM+S extension where a scalar of mass $m_S$ mixes has mixing angle  $\sin \theta$ with the 125 GeV Higgs boson.
The red contour is the sensitivity to $B$-meson decays. The different blue-purple contours illustrate the minimum BR$(h\to ss)$ value to which MATHUSLA would be sensitive. 
The projected constraint contour for the SHiP experiment 
is shown by the dashed orange contour. 
(b) Right-Handed Neutrino LLP simplified model where the mixing with active neutrinos is dominated by the 2nd generation. 
}
\label{f.MATHUSLASMSsensitivity}
\end{figure}

The MATHUSLA white paper~\cite{Curtin:2018mvb} discussed the primary physics case in great detail. Three main conclusions form the core motivation for constructing MATHUSLA:
\begin{itemize}\itemsep=1mm
\item[1.] \textbf{LLPs are fundamentally motivated.}

They could explain the Hierarchy Problem, Dark Matter, Neutrino Masses and the Baryon Asymmetry of the Universe~\cite{Curtin:2018mvb}. 
See Fig.~\ref{f.theorysummary} for a summary of these top-down motivations. 
Long lifetimes are also ubiquitous in the SM, providing bottom-up motivation for LLPs as a generic BSM signature independent of any particular theory bias.

\item[2.] \textbf{The LHC Main Detectors are blind to large regions of the LLP signature space.}

For example, searches for LLPs decaying to hadrons (leptons) with less than a few 100 GeV ($\sim$ 10 GeV) of visible energy in the event have particularly low trigger efficiency and are highly constrained by QCD and other backgrounds. 

\item[3.] \textbf{MATHUSLA reclaims sensitivity in these blind spots with orders of magnitude greater cross section/lifetime reach than the LHC main detectors, and the ability to discovery LLPs with masses ranging from MeV to TeV.}

This can be demonstrated with a few representative and well-motivated examples. Fig.~\ref{f.MATHUSLAhiggssensitivity}~(a) compares the sensitivity of MATHUSLA to hadronically decaying \emph{LLPs produced in exotic Higgs decays}  to the projected sensitivity of an ATLAS search in the muon system. \emph{MATHUSLA can probe three orders of magnitude smaller LLP production rates (or longer lifetime) than the LHC main detectors.} This is one of the most important LLP benchmarks, since it could provide the smoking gun for many BSM theories like Neutral Naturalness or general Hidden Valleys.

Figure~\ref{f.MATHUSLAhiggssensitivity}~(b) demonstrates that MATHUSLA can detect \emph{Long-lived Higgsinos with masses exceeding a TeV}, which arise in theories including gauge mediation, supersymmetric axion models, and R-parity violation. MATHUSLA is also able to probe \emph{low-mass LLPs} in the MeV-GeV range, including \emph{dark scalars or right-handed neutrinos}, via their production in exotic decays of $B$-hadrons, see Fig.~\ref{f.MATHUSLASMSsensitivity}.  

\end{itemize}

For LLP production cross sections in the pb range, \textbf{MATHUSLA can probe lifetimes approaching the $c\tau \lesssim 10^7$m upper limit from Cosmology~\cite{Fradette:2017sdd}. }
Achieving sensitivity to this cosmological limit of LLP parameter space for production rates corresponding to plausible exotic Higgs decay fractions is especially significant, since it would allow for MATHUSLA to verify the nature of any invisible Higgs decay signal observed at the HL-LHC main detectors: if MATHUSLA also sees a signal, we discover the LLPs that are produced in this invisible decay; if MATHUSLA sees no signal, it would strongly support the hypothesis that the invisible Higgs decay produced a cosmologically significant DM  candidate.

It is also important to note that \textbf{the sensitivity offered by MATHUSLA cannot be achieved using missing energy (MET) searches} at the main detectors. As demonstrated in~\cite{Curtin:2018mvb}, the mass reach of dedicated LLP searches, both at MATHUSLA and the main detectors, is much higher than the mass reach of MET searches for a large range of LLP lifetimes and production processes.

Finally, we comment on some important relationships between MATHUSLA and other experiments (beyond the existing LHC detectors) dedicated to the search for LLPs.

The high intensity of LHC collisions results in a large number of $B$-hadrons that can decay to light LLPs, allowing MATHUSLA to probe deep into the parameter space of low-mass LLP models. 
\textbf{This provides unique sensitivity that is highly complementary to that of proposed intensity-frontier experiments like SHiP~\cite{Lanfranchi:2243034,Anelli:2015pba,Alekhin:2015byh}.}

The Physics Beyond Collider (PBC) working group has carried out a detailed comparison of MATHUSLA, SHiP, FASER~\cite{Feng:2017uoz, Feng:2017vli} and CODEX-b~\cite{Gligorov:2017nwh} \textbf{limited to  minimal low-mass LLP models in the MeV-GeV range.} These comparisons are included in the PBC submission to this European Strategy Update.
While reach for minimal low-mass LLP models is important, MATHUSLA can take advantage of the full LHC collision energy to discover LLPS with weak- or TeV-scale masses. Furthermore, even low-mass LLPs can have additional production modes at the LHC that greatly increase sensitivity but are inaccessible at lower-energy experiments. An example are exotic Higgs decays, see Fig.~\ref{f.MATHUSLASMSsensitivity} (a).

FASER~\cite{Feng:2017uoz, Feng:2017vli} is a proposed small low-cost experiment to search for low-mass LLPs produced at very small angles to the LHC beam. 
Due to its shorter baseline, FASER is highly complementary to MATHUSLA and the two experiments probe almost mutually exclusive regions parameter space of minimal low-mass LLP models.
\textbf{This makes MATHUSLA and FASER in combination a particularly effective augmentation of the LHC's main detectors.}

\vspace{-3mm}
\subsection{Secondary Physics Goal: Study of Cosmic Rays}
\vspace{-3mm}

The design of MATHUSLA is driven by the robust tracking requirements of distinguishing upwards-traveling charged particles from LLP decays from downward-traveling cosmic rays (CR).
 It is therefore not surprising that MATHUSLA  can also act as a cutting-edge cosmic ray telescope. 
 The qualitative CR physics case was discussed in \cite{Curtin:2018mvb}. More detailed studies and a dedicated CR whitepaper are in progress.

MATHUSLA's $\sim (100 \mathrm{m})^2$ area gives it good efficiency for extended air showers (EAS) arising from primary cosmic rays with energies in the $\sim 10^{14} - 10^{18}$ eV range. This is a crucial energy window because it contains the main features in the energy spectrum of Galactic CRs, the so-called \emph{"knee"} at about 3--4 PeV and a second knee at about 100 PeV. The knee is connected with the end of the Galactic CR spectrum and the transition from Galactic to extra-galactic CRs. 
\textbf{A detailed study of the primary CR spectrum in this energy region is of great importance in astro-particle physics.} 

 Below the so-called ``knee'' at $E_{pr} \sim 10^{15} - 10^{16}$ eV, the spectrum is dominated by galactic sources like supernovae. The precise shape of the energy spectrum, its anisotropy, and its elemental composition (H, He, Fe, \ldots) provide crucial information on the  distribution of these sources and the astrophysics that drives them.
 At the upper range of MATHUSLA's possible sensitivity, $E_{pr} \sim 10^{17}-10^{18}$ eV, galactic CRs taper off and instead extragalactic sources dominate. This transition region is of great interest to study both the properties of the galactic magnetic field that confines charged particles within our milky way, as well as the characteristics of extragalactic sources and charged particle propagation in the intergalactic medium.

The combination of high-resolution directional tracking, near-full-area coverage, and proximity to ATLAS or CMS for correlated shower core measurements would allow MATHUSLA to make unique contributions to the study of this energy region. 
  Compared to KASKADE, which supplied measurements in this energy range, MATHUSLA's full area coverage with directional tracing allows it to measure energies with greater precision. The same feature also allows MATHUSLA to directly measure the CR flux at energies significantly below $10^{14}$ eV, which allows its measurements to be compared and calibrated to satellite measurements (e.g. CREAM, Calet, HERD).
  Compared to ARGO, which has similarly full detector coverage but is smaller in size with just a single detector layer, MATHUSLA has more tracking layers and benefits from proximity with the HL-LHC main detector. Its lower elevation than ARGO is also useful for calibrating systematic uncertainties of the observation. 
   Measurements that combine MATHUSLA and main detector data would provide unique insight into the dense inner core of extended air shower, offering the potential to probe CR primary composition at higher energies than KASKADE.

Since our understanding of this part of the CR spectrum is sensitively constrained by the amount of available statistics, MATHUSLA's measurements would allow it to make important contributions in resolving several long-standing puzzles and inconsistencies between other CR experiments. 
 \textbf{These CR observations represent a guaranteed physics return on the investment of constructing the detector.}
 
 \newpage



\bibliography{mathusla}
\bibliographystyle{JHEP}


\end{document}